\documentstyle[preprint,aps]{revtex}

\begin{document}
\title{The Relativistic Transport Model Description of Subthreshold Kaon
Production in Heavy-Ion Collisions}
\author{X. S. Fang, C. M. Ko, G. Q. Li,
and Y. M. Zheng\footnote{Permanent address:
Institute of Atomic Energy, Beijing 102413, China}}
\address{Cyclotron Institute and Physics Department\\
Texas A\&M University, College Station, Texas 77843}

\maketitle

\begin{abstract}
The relativistic transport model, in which the nucleon effective mass is
connected to the scalar field while its energy is shifted by the vector
potential, is extended to include the kaon degree of freedom.  We
further take into account the medium modification of the kaon mass
due to the explicit chiral symmetry breaking.
Both the propagation of kaons in the mean-field potential
and the kaon-baryon elastic scattering
are explicitly treated in our study.
We find that the attractive kaon scalar mean-field potential
in the dense matter leads to an enhanced kaon yield
in heavy-ion collisions at energies of about 1 GeV/nucleon
that are below the threshold for kaon production from the
nucleon-nucleon interaction
in free space. The final-state kaon-baryon scattering is seen to
affect significantly the kaon momentum spectra,
leading to an enhanced yield of kaons with large momenta or at large
laboratory angles.
With a soft nuclear equation of state and including the attractive
kaon scalar potential,
the calculated kaon energy spectra agree with
the preliminary data on subthreshold kaon production
from the heavy-ion synchrotron at GSI.
\end{abstract}

{\bf 1. Introduction}

\bigskip

Kaons were first measured in heavy-ion collisions at Bevalac in the 1980s
at incident energies around 2 GeV/nucleon \cite{NAGA81,LBL}.  These
experiments had generated many theoretical studies based on both the cascade
\cite{RK80,CUG84,SZ87} and the transport \cite{MOS90,LI92} model.
Recently, new experiments on
kaon production in heavy-ion collisions with both light and heavy
systems are being carried out
at the heavy-ion  synchrotron (SIS) at GSI \cite{GSI}.
The incident energy per nucleon in these experiments is
around 1 GeV and is below the threshold energy of 1.56 GeV for kaon
production in the nucleon-nucleon interaction in free space.
The study of kaon production in heavy-ion collisions at subthreshold
energies is interesting as it offers the possibility to investigate
the properties of the dense matter formed in the collisions through their
effects on the production mechanism
\cite{AICH85,AICH87,KO89}.
These effects include:

 (1) The intrinsic Fermi momentum of nucleons. It adds to the
incident momentum of the projectile nucleons and increases the
available energy for kaon production.
The nucleon Fermi momentum is relatively well determined
and is included in all transport models for  heavy-ion collisions.
However, at incident energies of about 1 GeV/nucleon, the available energy
after including the Fermi momentum
is still largely below the kaon production threshold.
Other medium effects are therefore needed to account for
the observed kaons in heavy-ion collisions at these energies.

 (2) Two-step processes involving first the excitation of a nucleon to
a resonance and then the production of a kaon from the
interaction of the resonance with other baryons.
At incident energies of around 1 GeV/nucleon,
only the delta resonance is appreciably produced.
The production and decay of the delta is usually included
in the transport model description of heavy-ion collisions at energies
above the pion production threshold. As shown in refs.
\cite{AICH85,HUAN93,AICH93}
the delta resonance indeed plays an important role in  subthreshold
kaon production.

 (3) The compressional energy of the nuclear matter.
It is clear that the smaller
is the nuclear matter compressional energy, the more is the incident energy
available to  kaon production and hence the larger is the kaon
yield. Actually, the determination of the nuclear matter incompressibility
is one of the motivations for carrying out experiments on subthreshold
kaon production.
As first suggested by Aichelin and Ko \cite{AICH85},
the kaon yield at incident energies below 1 GeV/nucleon
can differ by a factor of 2-3 depending on the stiffness of
the nuclear equation of state. This sensitivity of the kaon yield to the
incompressibility of the nuclear matter was later confirmed in similar
calculations \cite{MOS90,LI92,HUAN93,AICH93}. However, the momentum dependence
of the nucleon mean-field potential reduces the kaon yield
and its sensitivity to the nuclear equation of
state \cite{LI92,AICH87}.
The determination of the nuclear matter incompressibility
has thus become more difficult than proposed in ref. \cite{AICH85}.

 (4) The medium modification of kaon properties. The attractive
kaon scalar potential due to explicit chiral symmetry breaking reduces
the kaon production threshold in a medium and has been shown to
lead to an
enhanced production of kaons in heavy-ion collisions \cite{KO93}.
The study of
kaon production in heavy-ion collisions thus offers the possibility of
determining the kaon properties in a dense matter, which is
a topic of great interests in both
nuclear physics and astrophysics.

 (5) Multiparticle processes in which more than two nucleons
are involved in kaon production. Although the possibility of
simultaneous collisions of three or more nucleons is  small \cite{BAT92},
kaon production from such collisions is expected to be increasingly
important as the incident energy decreases. On the other hand,
the multiparticle contribution may have already been partly included
in transport models through multi-step processes involving the
excitation of resonances
and subsequent nucleon-resonance
and resonance-resonance interactions.

 (6) Kaon propagation in the mean-field potential and
the kaon-baryon elastic scattering.
Although kaons are not reabsorbed by baryons
as a result of strangeness conservation,
their momentum and angular distributions are, however,
appreciably modified by these final-state interactions
\cite{RAN81,FA93}.

In this paper, we shall study in detail the medium effects on subthreshold
kaon production in the framework of the relativistic transport model
\cite{KO87}, which is reviewed in section 2. The property of a kaon in a
dense matter is described in
section 3. In section 4, the kaon production cross sections
from baryon-baryon interactions in both free
space and heavy-ion collisions are discussed,
while in section 5 the propagation and rescattering
of the kaon in a nuclear medium are described. In section 6, results on
kaon production in nucleus-nucleus collisions at 1 GeV/nucleon
are presented and compared with
the data from  SIS. Finally, a summary is given in section 7.
Preliminary results of our work on kaon production in a Au+Au collision
at 1 GeV/nucleon have been reported in ref. \cite{FA93A}.

\bigskip\bigskip

{\bf 2. The relativistic transport model}

\bigskip

{}From the quantum hadrodynamics \cite{QHD} in which the nuclear matter is
treated as a system of interacting baryons and mesons, one can derive a
transport equation for the phase space distribution function $f(x,{\bf p^{*}}%
)$ of baryons while treating the meson degrees of freedom
as virtual in the sense that they act only as the mediators of the
baryon-baryon interactions \cite{KO87,MAL88,NORE90,WA91}. In
the following, we shall give a brief description of the
relativistic transport model based on ref. \cite{KO87}.

Assuming that the nuclear matter is quasihomogeneous and that the interactions
are weak, the resulting transport equation then reads as
\begin{equation}
\label{vuu} \frac 1{p_0^{*}}\{[\partial _x^\mu -(\partial _x^\mu \Sigma
_v^\nu -\partial _x^\nu \Sigma _v^\mu )\partial _\nu ^{p^{*}}]p_\mu ^{*}+m^*
({\partial _x^\mu } m^{*})\partial _\mu ^{p^{*}}\}f(x,{\bf p}^{*})=\,I_c,
\end{equation}
where $I_c$ is the collision term. In the above, the nucleon effective mass $%
m^*$ and kinetic momentum $p_\mu^*$ are defined by
\begin{equation}
m^*=\,m+\Sigma_s,\qquad p_\mu^*=\,p_\mu+\Sigma_{v\mu},
\end{equation}
where the scalar ($\Sigma_s$) and the vector ($\Sigma_{v\mu}$)
self-energy of the nucleon are given by
\begin{equation}
\Sigma_s=-g_\sigma\langle\sigma\rangle ,
\qquad \Sigma_{v\mu}=-g_\omega\langle\omega_\mu\rangle .
\end{equation}
The coupling constant of the nucleon to the scalar ($\sigma$) and
the vector ($\omega$) meson are denoted,
respectively, by $g_\sigma$ and $g_\omega$.

In the mean-field approximation, the
expectation value of the scalar field in
the nuclear matter $\langle\sigma\rangle$
is related to the nuclear scalar
density $\rho_s$ by
\begin{equation}
{m_\sigma}^2\langle\sigma\rangle +\,b\langle\sigma\rangle ^2
+\,c\langle\sigma\rangle ^3\approx\,g_\sigma \rho_s,
\end{equation}
where $m_\sigma$=~550 MeV
is the scalar meson mass. The last two terms on the left
hand side of the above equation is due to the self-interaction of the scalar
meson, i.e.
\begin{equation}
U(\sigma)=\,b\sigma^3/3\,+\,c\sigma^4/4.
\end{equation}
The expectation value of the vector field in the nuclear matter
$\langle\omega_\mu\rangle$ is related to
the nuclear current density $\rho_\mu$ by
\begin{equation}
\langle\omega_\mu\rangle =(g_\omega/m_\omega^2)\rho_\mu,
\end{equation}
where $m_\omega$=~783 MeV is the vector meson mass.

In terms of the phase-space distribution function, the scalar and the current
density in the local-density approximation can be expressed, respectively,
as
\begin{equation}
\rho_s=\,\int \frac{d^3{\bf p}^*}{(2\pi)^3}f(x,{\bf p}^*)m^*/E^*,
\end{equation}
and
\begin{equation}
\rho_\mu=\,\int \frac{d^3{\bf p}^*}{(2\pi)^3}f(x,{\bf p}^*)p_\mu^*/E^*,
\end{equation}
with $E^*=({\bf p^*}^2+{m^*}^2)^{1/2}$.

We shall consider two parameter sets from ref. \cite{LI93}.  The first
set is given by
\begin{eqnarray}\label{para1}
C_\omega& =&(g_\omega/m_\omega)m_N=8.498,\qquad
C_\sigma =(g_\sigma/m_\sigma)m_N=11.27, \nonumber\\
B&=&b/(g_\sigma^3m_N)=-2.83\times 10^{-2},\qquad
C=c/g_\sigma^4=0.186,
\end{eqnarray}
which gives a nuclear binding energy of 15.96 MeV/nucleon
at a saturation density of
0.16 fm$^{-3}$, a nucleon effective mass of 0.83 $m_N$, and an
incompressibility of 380 MeV. The second set is given by
\begin{eqnarray}\label{para2}
C_\omega& =&(g_\omega/m_\omega)m_N=8.498,\qquad
C_\sigma =(g_\sigma/m_\sigma)m_N=13.95, \nonumber\\
B&=&b/(g_\sigma^3m_N)=1.99\times 10^{-2},\qquad
C=c/g_\sigma^4=-2.96\times 10^{-3},
\end{eqnarray}
corresponding to the same nucleon effective mass of 0.83 $m_N$ but a
smaller nuclear incompressibility of 200 MeV.  In fig. 1, we show
the energy per nucleon, ${\cal E}/A$, as a function of density for the two
sets of parameters.  The solid and the dashed curve correspond to the
nuclear incompressibility of 380 MeV (stiff) and 200 MeV (soft),
respectively. In fig. 2, the
density dependence of the nucleon effective
mass for the two parameter sets
is shown by the solid (stiff) and the dashed (soft) curve.
It is seen that in both cases the nucleon effective mass
decreases with increasing nuclear density, and the reduction is
larger for the soft equation of state than for the stiff one.

The relativistic transport
equation is solved by the method of pseudoparticles \cite{WO82}.
In this method, each nucleon is replaced by a collection of test particles,
and the one-body phase-space distribution function is given by the
distribution of these test particles in the phase space. To solve the Vlasov
equation, i.e. the left hand side of the transport equation, is then
equivalent to the solution of the following classical equations of motion
for all test particles,
\begin{equation}\label{motion}
\frac{d{\bf x}}{dt}=\,{\bf p}^*/E^*, \qquad
\frac{d{\bf p}}{dt}=\,-{\bf\nabla}_x
[E^*+(g_\omega/m_\omega)^2\rho _B],
\end{equation}
with $\rho _B$ being the nuclear matter baryon density.

The collision term $I_c$ in eq. (\ref{vuu}) has the form
\begin{eqnarray}
I_c&=&\,\int{{d{\bf p}_2^*}\over(2\pi)^3}\int{{d{\bf p}_3^*}\over(2\pi)^3}\
\int{d\Omega}v\frac{d\sigma}{d\Omega}
\delta^3({\bf p}^*+{\bf p}_2^*-{\bf p}_3^*-{\bf p}_4^*)\nonumber\\
&&\cdot \{f(x,{\bf p}_3^*)f(x,{\bf p}_4^*)[1-f(x,{\bf p}^*)]
[1-f(x,{\bf p}_2^*)]
-f(x,{\bf p}^*)f(x,{\bf p}_2^*)[1-f(x,{\bf p}_3^*)]\nonumber\\
&&\cdot [1-f(x,{\bf p}_4^*)]\},
\end{eqnarray}
where $v$ is the relative velocity between the colliding nucleons and $\frac{%
d\sigma}{d\Omega}$ is the nucleon-nucleon differential cross section.

As in normal Vlasov-Uehling-Uhlenbeck
model \cite{BE88}, the isospin-averaged cross sections
in free space \cite{CUG81} are used for the elastic ($NN\rightarrow
NN$) and the delta excitation ($NN\rightarrow N\Delta $) process.
The cross section for the inverse process
$N\Delta\to NN$ is determined from the detailed balance
relation \cite{dan}.
Both the nucleon-delta ($N\Delta\rightarrow
N\Delta$) and the delta-delta ($\Delta\Delta\rightarrow\Delta\Delta$ )
elastic collision are also allowed; their
cross sections are  assumed to be the same as that for
the nucleon-nucleon elastic
scattering at the same center-of-mass energy. The mean-field potential for
a delta is taken to be similar to the one for a nucleon.

When a delta is formed, its mass distribution is taken to be of the
Breit-Wigner form
\begin{eqnarray}
P(m_\Delta )={(\Gamma  (q)/2)^2\over (m_\Delta -m_0)^2+(\Gamma (q)/2)^2},
\end{eqnarray}
where $ m _0$~=~1.232 GeV and the momentum-dependent delta
width \cite{TOKI86},
$\Gamma (q)$, is
\begin{eqnarray}
\Gamma (q)={0.47q^3\over [1+0.6(q/m_\pi )^2]m_\pi ^2}.
\end{eqnarray}
In the above, $q$ is the pion momentum in the center-of-mass frame of the
delta and is related to the delta mass $m_\Delta$ by
\begin{eqnarray}
q={\sqrt {\big[ m_\Delta ^2-(m_N+m_\pi )^2\big] \big[
m_\Delta ^2-(m_N-m_\pi )^2\big]}
\over 2m_\Delta}.
\end{eqnarray}

The collision between two baryons is treated in the same way
as in  the cascade model
\cite{CUG81}. A collision occurs when the distance between them is less
than $\sqrt{\sigma /\pi}$ with $\sigma $ being the interaction cross
section of the two baryons. After the collision, the directions of
the momenta of the two particles change in a statistical
way according to the empirical angular
distribution. Collisions are allowed only among particles in the same
simulation but the mean nuclear density and current are computed with all
test particles in the ensemble.  We note that in determining the baryon-baryon
center-of-mass, that is needed in treating their collision, we use the
kinetic momentum $p^*$ and energy $E^*$ of the two particles.
The relativistic transport model described in the above
has also been extensively studied
by the Giessen group \cite{MOS93}.

For heavy-ion collisions at about 1 GeV/nucleon, the pion degree of freedom
is non-negligible. It is included in the relativistic transport model through
the delta decay, i.e.,
$\Delta\to\pi N$, according to the momentum-dependent delta width
given by eqs. (14) and (15). The inverse reaction $\pi N\to\Delta$
is also included to take into account the pion absorption. The details on
treating the production and interactions of pions
in the transport model can be found
in ref. \cite{XI90}. As in most transport models for pions \cite
{WOLF90,BAO91}, the pion mean-field potential is not included so they
propagate as free particles in the nuclear medium. We would like to
point out,
however, that  medium effects are important and may lead to
an enhanced yield of pions with low transverse energies as shown
recently in ref. \cite{XI93}.

The relativistic transport model can be easily generalized to include the
hyperon degrees of freedom. From the generalized Walecka model \cite{GM91},
the hyperon effective mass $m_Y^*$ and kinetic momentum $p_{Y\mu}^*$
are given by
\begin{eqnarray}
m_{Y}^*=m_{Y}-g_{\sigma YY}\langle \sigma \rangle ,~~
p_{\mu}=p_{Y\mu}-g_{\omega YY}\langle \omega _\mu \rangle .
\end{eqnarray}
According to the quark model, the coupling constants $g_{\sigma YY}$
and $g_{\omega YY}$ of a hyperon to the scalar and the vector field
are 2/3 of those of a nucleon.
The hyperon mean-field potential is thus weaker than that of the nucleon,
and this is consistent with the hyperon-nucleus
phenomenology \cite{DO88}.  In fig. 1, the density
dependence of the $\Lambda$ mass is shown by the solid and the dashed curve
for the stiff and the soft equation of state, respectively.
The density dependence of the $\Sigma$ mass is similar
to that of the $\Lambda$ mass.

\bigskip\bigskip

{\bf 3. Kaons in a dense matter}

\bigskip

The  properties of a kaon in a dense matter was first studied in ref.
\cite{KL86}
and further investigated in refs.
\cite{BRO87,TAT88,BRO88,RHO94} based on  the chiral Lagrangian.
Recently, the Nambu$-$Jona-Lasinio model has also been used
to study the in-medium kaon properties \cite{WEI92}.
It has been shown that the medium modification of kaon properties has
significant effects in
both neutron stars \cite{BRO92} and
kaon production from high-energy heavy-ion collisions \cite{KO91,FA93B}.

{}From the chiral Lagrangian, Kaplan and Nelson \cite{KL86} showed that
nucleons
act on kaons as an effective scalar field because of
explicit chiral symmetry breaking, i.e.,
\begin{eqnarray}
L _S\approx {\Sigma_{KN} \over f^2_K} \bar NN\bar KK,
\end{eqnarray}
where $f_K$ is the kaon decay constant and $\Sigma _{KN}$ is the $KN$ sigma
term. This gives rise to an attractive $s$-wave interaction for the kaon.
There is also a vector interaction in the chiral Lagrangian,
\begin{eqnarray}
L_V\approx -{3i\over 8f_K^2} N^\dagger N \bar K\stackrel{\leftrightarrow}{%
\partial}_t K,
\end{eqnarray}
which leads to a repulsive $s$-wave interaction for the kaon.

Combining the scalar and the vector interaction with the kaon mass term
in the Lagrangian, we obtain the kaon in-medium mass
\begin{eqnarray}
m^*_K \approx m_K\Bigg( 1-{\Sigma _{KN}\over f_K^2m^2_K}\rho _S+{3\over 4}
{1\over f^2_Km_K}\rho _B\Bigg) ^{1/2},
\end{eqnarray}
where $\rho _S$ and $\rho _B$ are the scalar and the vector density of
baryons, respectively. The kaon in-medium mass thus depends sensitively on
the magnitudes of $f_K$ and
$\Sigma _{KN}$ which is defined by
\begin{equation}
\Sigma_{KN}=\,{\frac{1}{2}}(m_u+m_s)\langle N|{\bar u}u+{\bar s}s|N\rangle,
\end{equation}
with $m_u$ and $m_s$ being, respectively, the up and strange quark masses.
While
$f_K$ has a value similar to the pion decay constant $f_\pi\approx 93$
MeV \cite{BKK92}, the value for $\Sigma _{KN}$ is not well
determined because of the
considerable uncertainty in the strangeness content of a nucleon
\cite{WVC90}. The lower and upper limits on $\Sigma _{KN}$
can be obtained by
taking $\langle N|{\bar s}s|N\rangle
=\langle N|{\bar u}u|N\rangle$ and  $\langle N|{\bar s}s|N\rangle =0$,
respectively.
With $m_s = 25 ~\frac{m_u + m_d}{2}$ and the pion-nucleon Sigma-term $%
\Sigma_{\pi N} = \frac{m_u + m_d}{2} \langle N|{\bar u}u + {\bar d}
d|N\rangle \simeq 46
\, {\rm MeV}$, we have $300 \, {\rm MeV \leq \Sigma_{KN} \leq 600\, MeV}$.
With $f_K\approx 93$ MeV and a conservative value of
$\Sigma _{KN}\approx 350$ MeV, we find that the kaon in-medium mass,
including both the scalar and the vector self-energy of the kaon, increases
slightly with density, as shown in fig. 3 by the solid curve. This is
consistent with the fact that the kaon-nucleon scattering length in free
space is negative and small.

In ref. \cite{BKK92} the kaon in-medium mass
is defined in terms of the scalar interaction
only, while the vector field is included in the shift of the kaon energy.
This is similar to the treatment of the scalar and the vector mean
field for the nucleon  in both the Walecka
model \cite{QHD} and the Dirac-Brueckner approach \cite{MACH89,LI92A}.
The kaon in-medium mass is then
\begin{eqnarray}
m^*_K\approx m_K\Bigg( 1-{\Sigma _{KN}\over f^2_Km^2_K}\rho _S\Bigg) ^{1/2}.
\end{eqnarray}
In this case, the kaon in-medium mass decreases with density, as shown
in fig. 3 by the dashed curve. At the critical
density $\rho _c\approx f_K^2m_K^2/\Sigma_{KN}$, the kaon in-medium mass
becomes
zero. However, its energy is still positive due to the repulsive vector
interaction. The kaon energy using both definitions for the kaon in-medium
mass turns out to be very similar, and the results for kaon
production in heavy-ion collisions therefore does not depend on which
definition of the kaon in-medium mass is adopted.

{}From the KFSR relation $m_\rho=\,2{\sqrt
2} f_Kg_\rho$ and the SU(3) relation $g_\omega=\,3g_\rho$,  the
kaon vector potential is just 1/3 of the
nucleon vector  potential \cite{BKK92}. As
the vector potential of the hyperon is 2/3 of that of the
nucleon, the vector potential energies
in the initial and the final state of the process $B_1B_2\rightarrow BYK^+$
cancel each other. The kaon vector potential therefore does not
play any role in kaon production from baryon-baryon interactions
in a medium.

The one-loop corrections to the kaon scalar interaction based on the
chiral Lagrangian have recently been calculated by Lee {\it et al.}
\cite{RHO94}. They have found that loop corrections are repulsive and
reduce the attraction from the scalar mean field by $\sim$ 10\%. This
is significantly smaller than the loop corrections estimated by Lutz
{\it et al.} \cite{WEI92} using the Nambu$-$Jona-Lasinio model in
which the kaon is treated as a coherent quark-antiquark excitation
of the QCD vacuum.  On the other hand, the dense matter formed in heavy-ion
collisions is highly excited, and loop corrections are thus expected
to be suppressed. Because of this possible
suppression and also the large uncertainties
in theoretical predictions, we will not include loop corrections
in the present work.

\bigskip\bigskip

{\bf 4. Kaon production from heavy-ion collisions}

\bigskip

At incident energies around 1 GeV/nucleon, the kaon is mainly produced
from the baryon-baryon interaction in association with a hyperon, i.e.,
$B_1B_2\rightarrow BYK^+$. Here $B_1$, $B_2$ and $B$ are either a nucleon
or a delta, while $Y$ represents either a $\Lambda$ or a $\Sigma$ hyperon.
The contribution from the pion-nucleon interaction via $\pi N\to YK^+$
is only about 25\% \cite{XION90}.
To calculate kaon production from heavy-ion collisions, we need
to know the elementary kaon production cross section from the baryon-baryon
interaction in both free
space and a  medium.

The kaon production cross section in the nucleon-nucleon interaction can
be obtained from the  proton-proton and the proton-neutron interaction.
Randrup and Ko \cite{RK80} analysed the available experimental data and
proposed the following parametrization for the isospin-averaged
kaon production cross section from the nucleon-nucleon interaction in free
space
\begin{equation}
\sigma _{NN\rightarrow BYK^+} (\sqrt s)=36 ~{p_{\rm max}\over m_K} ~\mu {\rm
b},
\end{equation}
where the kaon maximum momentum $p_{\rm max}$ is related to the nucleon-nucleon
center-of-mass energy $\sqrt s$ by
\begin{equation}
p_{\rm max}={1\over 2}\sqrt {\big[ s-(m_B+m_Y+m_K)^2\big]
\big[ s-(m_B+m_Y-m_K)^2\big] /s}.
\end{equation}
In fig. 4, we compare this parametrization (isospin factor corrected)
with the experimental data for $pp\rightarrow p\Lambda K^+$ \cite{com}.

The kaon production cross sections from the
nucleon-delta and the delta-delta interaction
were also analysed in ref. \cite{RK80} based mainly on the isospin arguments.
It was found that
\begin{equation}
\sigma _{N\Delta\rightarrow BYK^+}(\sqrt s)\approx {3\over 4}
\sigma _{NN\rightarrow BYK^+}(\sqrt s),
\end{equation}
\begin{equation}
\sigma _{\Delta\Delta\rightarrow BYK^+}(\sqrt s)\approx {1\over 2}
\sigma _{NN\rightarrow BYK^+}(\sqrt s).
\end{equation}

We note that in eq. (22), $p_{\rm max}$ depends on masses of the final-state
particles, and the kaon production cross section is thus different for
different final states, even if the initial state is the same. For
the reaction $NN\rightarrow BYK^+$, we show in fig. 5 the kaon production
cross section
corresponding to four possible final states: $N\Lambda K^+$, $N\Sigma K^+$,
$\Delta\Lambda K^+$ and $\Delta\Sigma K^+$.
In our calculation, all four possible channels are included, although
at incident energies considered here, contributions from
processes with a delta in the final state are insignificant.

In addition to the total kaon production cross section, we also need the kaon
momentum distribution from the baryon-baryon interaction in free space.
This has been
parametrized in ref. \cite{RK80} according to the phase space argument,
\begin{equation}
{E\over p^2}{d^3\sigma (\sqrt s)\over dpd\Omega}=
\sigma _{K^+}{E\over 4\pi p^2}{12\over p_{\rm max}}
\left (1-{p\over p_{\rm max}}\right )\left ({p\over p_{\rm max}}\right )^2,
\end{equation}
where $\sigma _{K^+}$ is given by eqs. (21), (23) or (24), while
$p_{\rm max}$ is given by eq. (22). This
parametrization has been shown to reproduce reasonably the experimental data
\cite{RK80}.

To determine the kaon production cross section from
the baryon-baryon interaction in a medium, we simply use
the in-medium masses  to calculate $p_{\rm max}$, i.e.,
\begin{equation}
p_{\rm max}={1\over 2}\sqrt {\big[ s^*-(m_B^*+m_Y^*+m_K^*)^2\big]
\big[ s^*-(m_B^*+m_Y^*-m_K^*)^2\big] /s^*}.
\end{equation}
where $m_Y^*$ is the hyperon in-medium mass, and $\sqrt {s^*}$ is the
center-of-mass energy of the two baryons available for kaon production.
Since the difference in the vector potential energies between baryons
in the initial and the final state of
$B_1B_2\rightarrow BYK^+$ is ${1\over 3}(g_\omega /m_\omega )^2$,
the center-of-mass energy $\sqrt {s^*}$ available for kaon
production is then given by $({m_{B_1}^*}^2+{{\bf p}_{B_1}^*}
^2)^{1/2}+({m_{B_2}^*}^2+{{\bf p}_{B_2}^*}^2)^{1/2}+{1\over 3}(g_\omega
/m_\omega)^2$.

The kaon production cross section in the baryon-baryon interaction
can also be determined theoretically.
In ref. \cite{KO89}, this was carried out  in the one-pion-exchange model
by including medium effects on the exchanged pion through the
delta-hole model. The medium modification of baryons, hyperons and
kaons was, however, neglected. Recently,
the one-kaon-exchange contribution to
kaon production in the proton-proton
interaction has been investigated \cite{LA91}.
It is shown that the contribution from the kaon exchange
is as important as that from the pion exchange if off-shell effects
are included, and both are needed
to account for the experimental data.
It will be of interest to continue the theoretical studies of the
elementary kaon production cross section,
as this would allow us not only to include the medium effects
but also to treat consistently
kaon production from the nucleon-nucleon, the nucleon-delta and the
delta-delta
interaction, as the latter two processes are experimentally inaccessible.

Because of the small probability for kaon production in the baryon-baryon
interaction, kaon production in heavy-ion collisions at subthreshold
energies is usually treated perturbatively so that the collision dynamics
is not
affected by the presence of the produced kaons. When the energy in a
baryon-baryon collision is above the threshold for kaon production,
the kaon is
produced isotropically in the baryon-baryon center-of-mass frame with a
momentum distribution taken to be the same as in free space given by eq. (25).

For a given impact parameter $b$ in a heavy-ion collision, the
Lorentz-invariant
double differential kaon multiplicity in the nucleus-nucleus center-of-mass
frame is  calculated from
\begin{equation}\label{kaon}
\frac E{p^2}\frac{d^2N(b)}{dpd\Omega }=\,\sum_{Coll.}\frac
1{\sigma _{total}} \frac{E^{\prime }}{p^{\prime }{}^2}\frac{d^2\sigma
({\sqrt{s}})}{dp^{\prime }d\Omega ^{\prime }}\int \frac{d\Omega }
{4\pi }[1-f({\bf r,p_B},t)].
\end{equation}
In the above, the summation is over all baryon-baryon collisions. The kaon
production cross section $\frac{E^{\prime }}{p^{\prime }{}^2}\frac{d^2\sigma
({\sqrt{s}})}{dp^{\prime }d\Omega ^{\prime }}$ is evaluated in the
baryon-baryon center-of-mass frame. The factor $[1-f({\bf r,p_B},t)]$ is the
available phase space for the final baryon with momentum ${\bf p_B}$ in the
nucleus-nucleus center-of-mass frame. The arbitrariness in the angle between
the nucleon and the
hyperon is taken into account by averaging over this angle. We
have ignored the phase space factor for the
hyperon as its abundance is negligibly small at incident energies
around 1 GeV/nucleon.
In the following calculation, we find that the Pauli-Blocking
effect in eq. (\ref{kaon}) reduces the kaon production probability by
less than 20\%.

The Lorentz-invariant double differential cross section for kaon production in
heavy-ion collisions is then obtained by summing over the impact parameter,
i.e.,
\begin{equation}\label{k3}
\frac{E}{p^2}\frac{d^2\sigma}{dpd\Omega}=\,2\pi\int bdb \frac{E}{p^2}\frac{%
d^2N(b)}{dpd\Omega}.
\end{equation}

\bigskip\bigskip

{\bf 5. Kaon propagation and rescattering in nuclear medium}

\bigskip

Because of the mean-field potential it feels
in the nuclear medium, the motion of a kaon
is similar to that for a nucleon. Representing the kaon by
test particles, its motion is then given by
equations similar to those in eq. (\ref{motion}), i.e.,
\begin{equation}
\frac{d{\bf x}_K}{dt}=\,{{\bf p}_k}^*/E_K^*, \qquad
\frac{d{\bf p}_K}{dt}=\,-{\bf\nabla}_x E_K^*,
\end{equation}
where the kaon energy is $E_K^{*}=({m_K^{*}}^2+{{\bf p}_K^{*}}^2)^{1/2}$
with $m_K^*$ given by eq. (18).

As the number of kaons produced in each event is very small,
their interactions with the nuclear
medium has usually been neglected in previous studies.
We have recently included both the kaon propagation and rescattering
in the relativistic transport model by
artificially producing $N_K$ kaons in
each baryon-baryon collision
above the kaon production threshold \cite{FA93,FA93A}.
We then associate each kaon
with a production probability, which is given by the ratio of the kaon
production cross section to the baryon-baryon total cross section. The
motions of these kaons are
followed by solving the classical equations of
motion as other particles in the model.

As in the treatment of the baryon-baryon collision,
a kaon collides with a nucleon if the
distance between them is less than $\sqrt{\sigma _{KN}/\pi} $, where $%
\sigma _{KN}$ is the kaon-nucleon total cross section and has a value of
about 10 mb \cite{DW82}, which we take to be density-independent. After
the collision, the direction of the kaon momentum is taken
to be isotropically distributed as the
kaon-nucleon interaction is mainly  $s$-wave. Since kaon production is
treated perturbatively, its effect on the nucleon dynamics is neglected. We
therefore do not allow the nucleon momentum to change in the kaon-nucleon
interaction. We have found \cite{FA93} that kaons undergo substantial
rescatterings as they are mostly produced in the
high density region of heavy-ion collisions. This leads
to significant effects on the final kaon kinetic energy spectra.

Similarly, we include also the collision between
a kaon and a pion via the $K^*$
resonance \cite{KO81}. Its effect on kaons produced
in heavy-ion collisions at subthreshold energies
is, however, insignificant compared with the kaon-nucleon interaction. This
is due to both the smaller number of pions than nucleons in the system and
the fact that pions materialize from delta decays in the expansion stage of
heavy-ion collisions when the matter density starts to decrease.

To obtain the final kaon spectrum, we weight each kaon with the production
probability introduced previously and then divide the sum by $N_K$.
Usually we choose $N_K$ to be $\sim$ 50 for light systems and $\sim$
25 for heavy systems.

\bigskip\bigskip

{\bf 6. Results and discussions}

\bigskip

We have carried out calculations
for kaon production in heavy-ion collisions at 1 GeV/nucleon for both
the light (Ne+Ne) and the heavy (Au+Au) system. The
results shown in this section are, unless otherwise explicitly
stated, obtained with the soft equation of state
and including all the medium effects discussed above.

In the calculations, the number of test particles used are
50 for Au+Au collisions and 500 for Ne+Ne collisions. With the
improved method of treating kaon production described in the last
section, the number of kaons in each momentum bin is at least more than
a hundred for low momenta and a thousand for high momenta.
We have thus sufficient statistics to show the theoretical kaon spectra
in the following figures by smooth lines.

In fig. 6, we show the time evolution of the baryon (nucleon
and delta), the delta, and
the pion density in the central region of
a head-on ($b$=0 fm) Au+Au collision at 1 GeV/nucleon.
It is seen that for a time interval of about 15 fm/c, a piece of
dense matter with a baryon density of 2-3$\rho _0$ is formed in the central
region (a cubic of 27 fm$^{-3}$) of the colliding system. Of all baryons
in this region, about 15\% are delta resonances with a peak density
of about 0.5$\rho _0$, which later decay into pions.
In fig. 7 the time evolution of the
abundance of deltas, pions and kaons  are shown.
Although both the delta and the pion number vary with time, their sum remain
essentially constant after the system reaches the maximum density.
Kaons, with an abundance of about
0.4, are mainly produced at the early stage of the collision
when the baryon-baryon collisions are most energetic and the density
is high. The kaon abundance reaches its maximum value
at about 15 fm/c when the system starts to expand.

In fig. 8, we separate the total kaon production probability of fig. 7 into
contributions from the nucleon-nucleon, the nucleon-delta and the delta-delta
interaction. It is seen that the contribution from the
nucleon-delta interaction accounts
for more than half of the total kaon yield, while the sum
of the contributions from the  nucleon-delta and the delta-delta interaction
is about 75\% of the total kaon yield.
It is thus important to
treat correctly the delta dynamics in heavy-ion collisions.
In our calculation, the contribution from the delta-delta interaction
turns out to be slightly smaller than that from the
nucleon-nucleon interaction,
while in refs. \cite{HUAN93,AICH93}, it has been shown that the
contribution from the delta-delta interaction is more important than
that from the  nucleon-nucleon interaction. The difference between our
results and those of refs. \cite{HUAN93,AICH93} is mainly due to the following
two reasons. Firstly, in the calculations of refs. \cite{HUAN93,AICH93},
deltas are frozen during the dynamical evolution of the colliding system
and allowed to decay only at the final stage of the collision. This
treatment apparently overestimates the contributions from the  nucleon-delta
and the delta-delta interaction. Secondly, kaon production
in refs. \cite{HUAN93,AICH93}
was calculated in the framework of the quantum
molecular dynamics with a Skyrme-type momentum-independent mean-field
potential. As is well-known, the momentum-dependence of the nucleon
mean-field potential, which is automatically included in our
relativistic transport model, reduces the number of two-body collisions
and hence the delta abundance, thus leading to a smaller contribution
from the delta-delta interaction as compared to the case with the
momentum-independent mean-field potential.
Indeed, in a test calculation with the non-relativistic
Vlasov-Uehling-Uhlenbeck approach using the momentum-independent Skyrme-type
mean-field potential, we also find that the contribution from
the delta-delta contribution is
larger than that from the nucleon-nucleon interaction.

The effect of the nuclear equation of state on the kaon yield
is shown in fig. 9. The
solid and the dashed curve correspond to the
results obtained with the stiff (eq. (9))
and the  soft (eq. (10)) equation of state, respectively. The
kaon yield with the soft equation of state is about 15\% larger than that
with the stiff one. This difference is smaller than the prediction from
non-relativistic transport model calculations \cite{AICH85,HUAN93,AICH93}.
For example, in ref. \cite{AICH85} it was shown that at 1 GeV/nucleon,
the kaon yield with a soft equation of state is about 50\% larger than
that with a stiff one, which is confirmed in our test calculation
based on the non-relativistic Vlasov-Uehling-Uhlenbeck equation with the
momentum-independent Skyrme-type mean-field potential.
The less sensitivity of the kaon yield to the nuclear equation of state
in the relativistic transport model is mainly due to the fact that
the difference between the maximum densities
reached in the collision using the relativistic equations of state
is smaller than
that using the Skyrme-type equations of state.
On the other hand, as shown in ref. \cite{AICH85}, the difference between
the kaon yields with different equations of state increases with decreasing
incident energy. This is probably also true for the
relativistic transport model.
It is therefore desirable to carry out both theoretical and
experimental studies of subthreshold kaon production at incident energies
around 600-800 MeV/nucleon.

To see whether we have treated correctly the baryon, especially the delta,
dynamics, we compare, in fig. 10,
the pion momentum spectra from our calculations with recent
experimental data \cite{GSI} for a Au+Au collision at 1 GeV/nucleon.
The theoretical results are shown for four laboratory angles, while
the experimental data are for $\theta _{\rm lab}=44^o$. Indeed,
our theoretical results are in good agreement with the experimental data,
except that the theoretical spectrum
is slightly steeper than the experimental one.
This may be due to the medium effects on pions
\cite{XI93} which have not been included in  present calculations.

The comparison of our results for the kaon momentum spectra with the recent
experimental data from  SIS \cite{GSI} is shown in fig. 11 for a
Ne+NaF collision at 1 GeV/nucleon and in fig. 12 for a Au+Au collision at
1 GeV/nucleon. Again, the theoretical results are shown for four laboratory
angles, while the experimental data are for $\theta _{\rm lab}$=~44$^o$. We
see that the theoretical results with the soft equation of state and
including the kaon medium effects are in good agreement with the experimental
data for both colliding systems. We note that the results with the stiff
equation of state are essentially the same.

Reasonable agreements with the experimental data
were also obtained in refs. \cite{HUAN93,AICH93}
in the framework of the quantum molecular dynamics with a
Skyrme-type
nucleon mean-field potential.
In these calculations, the momentum dependence
of the nucleon mean-field potential is neglected. Furthermore,
deltas are allowed to decay only at the final stage
of the collision,  and the threshold for
$B_1B_2\rightarrow B\Sigma  K^+$, which is slightly higher (see fig. 5), is
assumed to be the same as that for $B_1B_2\rightarrow B\Lambda K^+$.
If these simplifications are corrected, results of refs. \cite{HUAN93,AICH93}
will be about a factor of 3-4 below
the experimental data. This discrepancy can be removed by including
the kaon
medium effects, especially the attractive kaon scalar mean-field
potential (see discussions
below).

In fig. 13, we show the total kaon production cross section in heavy-ion
collisions at 1 GeV/nucleon for a number of colliding systems.
The open squares are our
results based on the soft equation of state and including the kaon medium
effects. The open and the solid circle with error bars are the experimental
data \cite{GSI} for the
Ne+NaF and the Au+Au collision at 1 GeV/nucleon, respectively.
It is seen that our results for the total kaon production cross section
are in very good agreement with the experimental data. The variation of
the kaon production cross section with the atomic number of the projectile
nucleus can be expressed by a power law, i.e., $\sigma _{K^+}\sim A^{\tau}$
with $\tau\approx 2.2$.  A similar $A$ dependence of the kaon yield in
heavy-ion collisions has been previously pointed out in ref. \cite{RK80}.

We have also carried out a calculation in which the
kaon scalar mean-field potential is neglected. The results
for kaon production in a Au+Au collision at 1 GeV/nucleon
are shown in fig. 14. The solid
and the  dashed curve correspond to results with  and without the
kaon scalar mean-field potential, respectively. It is seen that without the
attractive kaon scalar mean-field potential,
the theoretical results are about a factor of 3-4
below the experimental data.
This is consistent with the results from the
non-relativistic transport model
calculations with the momentum-dependent nucleon mean-field
potential \cite{AICH93}. We note that in the relativistic transport model
this momentum dependence is automatically included, as the nucleon
Schr\"odinger-equivalent potential in a relativistic approach
is momentum dependent \cite{LI93A}.
The attractive kaon scalar potential reduces the kaon production
threshold in the medium and thus increases its yield.

In ref. \cite{maru93}, subthreshold kaon production from heavy-ion collisions
has also been studied in the relativistic transport model. In their
calculations, the medium effects on kaons are not included so kaons
are treated as free particles.  Their results also agree with the
experimental data and are thus comparable to ours.  Although there is
no mean-field potential for the kaon in ref. \cite{maru93},
the scalar
and vector potential for the hyperon are taken to have the same strength
as those for the nucleon, which is stronger than the hyperon mean-field
potential used in our calculations, i.e., 2/3 of the nucleon
mean-field potential.
In ref. \cite{maru93}, the threshold for kaon production in the
medium is thus also reduced as a result of the larger reduction in the
hyperon mass than in our work.  It is thus not surprising
that their kaon yields obtained without the kaon medium effects
agree also with the experimental data.

In fig. 15, we discuss the effect of final-state interactions
on the kaon momentum spectra. The theoretical kaon momentum spectra obtained
with (solid curves) and without
(dashed curves) the kaon final-state interactions (propagation in
the mean-field potential and rescattering)
are compared with the experimental data.
We see that the  final-state interactions modify the kaon momentum
spectra considerably, leading to an increase of both the kaon yield and the
slope of its spectra at large angles.
At $\theta _{\rm lab}=~44^o$, the kaon final-state interactions help bring the
theoretical results in better agreement with the experimental data.

In fig. 16, the distribution of the number of kaon rescattering is shown
for the Ne+Ne (dashed curve) and the Au+Au (solid curve)
head-on collision at 1 GeV/nucleon.
We find that on the average each
kaon undergoes about 2 rescatterings for the Ne+Ne collision
and 4  rescatterings for the Au+Au collision, which are similar to our recent
work for Ca+Ca collisions \cite{FA93} but much larger than that
of ref. \cite{RAN81} based on a schematic estimate. From fig. 15 we
conclude that, although kaons will not be absorbed once they are produced in
a hadronic matter, the elastic kaon-baryon rescattering modifies
appreciably their
momentum spectrum  and should be taken into account explicitly in
the transport model.

\bigskip\bigskip

{\bf 7. Summary and outlook}

\bigskip

We have generalized the relativistic transport model to include the kaon
mean-field potential and the collisions of kaons with other particles. Both
the scalar and the vector mean-field potential are obtained from the chiral
Lagrangian. In a nuclear medium,
nucleons act on the kaon as an effective scalar field, leading to an
attractive kaon scalar potential.
The kaon also feels a repulsive
vector potential with a strength about 1/3 of that for the  nucleon.
The attractive scalar potential reduces the
threshold for kaon production from the baryon-baryon
interaction in the medium,
which is the dominant process for its production in heavy-ion
collisions at subthreshold energies.
The kaon yield is thus enhanced by the attractive scalar potential.

The repulsive vector potential acting on kaons affects their kinetic energy
distribution. On their way out of the nuclear matter, kaons are accelerated by
the mean-field vector potential and acquire thus
more kinetic energy. However, this effect is somewhat reduced by the
final-state rescatterings of kaons by nucleons.
We find that the kaon undergoes
substantial number of elastic rescatterings. This leads to not only an
increase in the number of high energy kaons but also an enhancement of the kaon
yield at large angles.

Our results on the kaon momentum  spectra in both the Ne+NaF and the Au+Au
collision  at 1 GeV/nucleon are in good
agreement  with the recent experimental data from the SIS at
GSI \cite{GSI} when the attractive kaon scalar mean-field
potential and the kaon final-state
interactions are included. Neglecting the kaon scalar mean-field
potential reduces the kaon yield by about a factor of 3.

We have not found a significant difference between the kaon yields
obtained with different nuclear equations of state.
This is mainly due to the fact that the difference between the maximum
densities reached in the collision is smaller for the
different relativistic equations of
state than that for the different Skyrme-type equations of state.
We expect, however,
that the effect of the nuclear equation
of state on the kaon yield will be appreciable  in heavy-ion collisions
at incident energies below 1 GeV/nucleon.

Because of the lack of data on kaon production from the nucleon-nucleon
interaction
near the threshold and the complexity of the heavy-ion
collision dynamics,
we do not consider our results to be final. Our conclusion that
the kaon has a strong attractive scalar mean-field potential
in the dense matter is thus
tentative but very encouraging.
With more refined studies in the future,
we believe that we can learn from subthreshold kaon production
in heavy-ion collisions more about the nuclear equation of state at
high densities and the properties of a kaon in a dense matter.

\bigskip

We thank Gerry Brown and
Ulrich Mosel for useful conversations.
Also, helpful discussions with Eckard Grosse on the experimental data
are gratefully appreciated.
This work was supported in part by the NSF Grant No. PHY-9212209 and the
Welch
Foundation Grant No. A-1110.

\vfill\eject

\vfill\eject

\noindent{\bf Figure Captions}

\bigskip

\itemitem{Fig. 1.} The density dependence of the
energy per nucleon in nuclear matter.
The solid and the dashed curve
correspond to the stiff and the soft equation of state,
respectively. $\rho _0$=0.16 fm$^{-3}$.

\itemitem{Fig. 2.} The density dependence of the nucleon (N)
and the lambda ($\Lambda$) in-medium
mass. The solid and dashed curve correspond to the  stiff and the soft
equation of state, respectively.

\itemitem{Fig. 3.} The density dependence of the kaon in-medium mass.
The solid curve includes both the scalar and the vector mean-field potential,
while the dashed curve includes only the scalar mean-field potential.

\itemitem{Fig. 4.} The comparison of the parametrization, eq. (22) (with
isospin factor corrected), with the experimental data for $pp\rightarrow
p\Lambda K^+$ \cite{com}.

\itemitem{Fig. 5.} The kaon production cross section
for different final states in a nucleon-nucleon
interaction.

\itemitem{Fig. 6.} The time evolution of the baryon ($N+\Delta$),
the delta ($\Delta$), and the pion ($\pi $) density
in the central region of a head-on Au+Au collision at 1 GeV/nucleon.

\itemitem{Fig. 7.} The time evolution of the delta ($\Delta$),
the pion ($\pi$), and the kaon ($K^+$)
abundance in a head-on Au+Au collision at 1 GeV/nucleon.

\itemitem{Fig. 8.} Contributions to the total kaon production probability
from the nucleon-nucleon ($NN$), the nucleon-delta
($N\Delta$) and the delta-delta ($\Delta\Delta$)
interaction in a head-on Au+Au collision at 1 GeV/nucleon.

\itemitem{Fig. 9} The time evolution of the kaon production probability.
The solid and the dashed curve correspond to the results obtained with
the stiff and the soft equation of state, respectively.

\itemitem{Fig. 10.} The comparison of the theoretical pion momentum spectra
with the experimental data \cite{GSI} for a Au+Au collision at 1 GeV/nucleon.

\itemitem{Fig. 11.} The kaon  momentum  spectra at different laboratory
angles in a Ne+NaF collision at 1 GeV/nucleon.
The experimental data \cite{GSI} at 44$^0$
are also shown.

\itemitem{Fig. 12.} Same as Fig. 11 for a Au+Au collision at 1 GeV/nucleon.

\itemitem{Fig. 13.}  The total kaon production cross section as a function
of the atomic number of the projectile nucleus. The open squares are
the theoretical results, while the open and the solid circle are
experimental data \cite{GSI} for the Ne+NaF and the Au+Au collision,
respectively.

\itemitem{Fig. 14.} The kaon momentum spectra in a Au+Au collision at 1
GeV/nucleon, obtained with (solid) and without (dashed) the kaon scalar
potential,
respectively. The experimental data \cite{GSI} at $44^o$ are also
shown.

\itemitem{Fig. 15.} The kaon momentum spectra in a Au+Au
collision at 1 GeV/nucleon, obtained with (kaon) and without (dashed)
kaon final-state interactions (propagation in the mean-field potential and
rescattering). The experimental data \cite{GSI} at $44^o$ are also shown.

\itemitem{Fig. 16.} The normalized
distribution of the number of kaon rescatterings
in the Ne+Ne (dashed curve) and the
Au+Au (solid curve) head-on  collision at 1 GeV/nucleon.

\end{document}